\documentclass [12pt]{article}
\usepackage{amsmath}
\usepackage{graphicx}
\RequirePackage{comment}
\excludecomment{frontmatter}

\begin{document}

\begin{frontmatter}

\title{Comparison of Spline with Kriging in an Epidemiological Problem}

\begin{aug}
\author{\fnms{Roshanak} \snm{Alimohammadi}\corref{}\ead[label=e1]{r_alimohammadi@alzahra.ac.ir}}
\address{Department of Mathematics,
 Alzahra University, Tehran, Iran}
\affiliation{Alzahra University}
\end{aug}

\begin{abstract}
There are various methods to analyze different kinds of data sets.
Spatial data is defined when data is dependent on each other based
on their respective locations. Spline and Kriging are two methods
for interpolating and predicting spatial data. Under certain
conditions, these methods
 are equivalent, but in practice they show different behaviors.

Amount of data can be observed only at some positions that are
chosen as positions of sample points, therefore, prediction of
data values in other positions is important. In this paper, the
link between Spline and Kriging methods is described, then for an
epidemiological two dimensional real data set, data is observed
in geological longitude and in latitude dimensions, and behavior
of these methods are investigated. Comparison of these
performances show that for this data set, Kriging method has a
better performance than Spline method.
\end{abstract}


\begin{keyword}
\kwd{Spatial data}\kwd{Spline}\kwd{Kriging}
\end{keyword}

\end{frontmatter}

\title{Comparison of Spline with Kriging in an Epidemiological Problem}

\author{
{\bf   Roshanak Alimohammadi }\\
\ {\normalsize Department of Mathematics},\\
\normalsize Alzahra University,\\ Tehran, Iran.}
\date{ }
\maketitle
\begin{abstract}
There are various methods to analyze different kinds of data sets.
Spatial data is defined when data is dependent on each other based
on their respective locations. Spline and Kriging are two methods
for interpolating and predicting spatial data. Under certain
conditions, these methods
 are equivalent, but in practice they show different behaviors.

Amount of data can be observed only at some positions that are
chosen as positions of sample points, therefore, prediction of
data values in other positions is important. In this paper, the
link between Spline and Kriging methods is described, then for an
epidemiological two dimensional real data set, data is observed
in geological longitude and in latitude dimensions, and behavior
of these methods are investigated. Comparison of these
performances show that for this data set, Kriging method has a
better performance than Spline method.
\\
{\it {\bf Key Words:} Spatial data, Spline, Kriging.}
\end{abstract}
\section {Introduction}
To analyze every kind of data, a
 model of data structure can be
considered. In spatial data analysis, a random field $\{Z(t), t
\in D \subset R^d\}$ is applied for spatial data modeling, where
$\it{t}$
 is the
site of desired location and $\it{D}$ is an index set. For each t,
the random field $\it{Z(t)}$ can be decomposed as
\begin{eqnarray}
\label{1}
Z(t)= \mu(t) + \delta(t)
\end{eqnarray}
where $\mu(t)$ is the trend of random field and $\delta(t)$ is a
zero mean random field. In spatial statistics, there are many
methods for predicting the value of random field at a given
spatial site, say $\it{t_0}$, using observations ${\bf Z} =
(Z(t_1), \ldots , Z(t_n))'$ of the random field $Z(.)$ at $\it{n}$
spatial sites $t = (t_1,  \ldots ,t_n)$. One of the methods,
named Kriging, is the best unbiased predictor which has different
kinds such as Ordinary and Universal Krigings. In ordinary
kriging, the trend term in relation (\ref{1}) is fixed and in
universal kriging $\mu(t)$ is a function of $\it{t}$ (cressie
(1993)).

Spline is another method for spatial data prediction which
minimizes penalized sum of squares criterion. For more details
about Splines can refer to Green and Silverman (1994), Hart (2005)
and Hardle (2006).

Some authors studied the link between Spline and Kriging, as two
prediction methods. Theoretical link between these methods is
studied by Kent and Mardia (1994) and the applied link for some
data sets is investigated by Hutchinson and Gessler (1994) and
Lasslet (1994).
In this paper these methods are applied for predicting values of
data with two dimensional positions. This data set relates to
taberculosis infection prevalence in some cities of Iran which
observed in geological longitude and in latitude dimensions. A
brief review of Kriging and Spline methods are given respectively
in sections 2 and 3, and in section 4 these methods are applied
for predicting rate of taberculosis infection prevalence, and
performances of the methods are compared. Finally results and
conclusions are given in section 5 and the better method to
predict values of taberculosis infection
prevalence is determined based on the data. 
\section{Kriging}
In Universal Kriging, the trend term in relation (\ref{1})
 is an unknown linear combination of known
functions $f_j(.)$ with unknown coefficients $\beta_j$, that is
\begin{eqnarray*}
\mu(t) = \Sigma_{j=1}^{P+1}\beta_{j-1}f_{j-1}(t)
\end{eqnarray*}
where $\beta = (\beta_0, ... ,\beta_p)^{\prime}\in R^{p+1}$, is
an unknown vector of parameters. Furthermore, data {\bf Z}  can
be written as
\begin{eqnarray*}
{\bf Z}= X \beta + \delta
\end{eqnarray*}
where X is an $n \times (p+1)$ matrix whose (i,j)th element is
$f_{j-1}(t_i)$.

It is desired to predict $Z(t_0)$ linearly from data {\bf Z},
that is
\begin{eqnarray}
\label{3} \hat Z (t_0) = \lambda^{\prime} {\bf Z}, ~~~~~~~~\lambda
^{\prime} X = x^{\prime}
\end{eqnarray}
which is uniformly unbiased ($E[\hat Z (t_0)] = E[Z(t_0)]$), and
minimizes the mean squares error term $\sigma^2_e = E[(\hat Z
(t_0) - Z(t_0))^2]$ over  $\lambda=(\lambda_1, ..., \lambda_n)$.

Assumption $\lambda ^{\prime} X = x^{\prime}$ in equation
(\ref{3}) is equivalent to uniformly unbiased condition, where
$x= (f_0(t_0),...,f_n(t_0))^{\prime}$. Then the optimal value of
$\lambda$ in relation (\ref{3}) is
\begin{eqnarray}
\label{5}
\lambda^{\prime} =[C+X(X'\Sigma^{-1} X)^{-1}(x - X'\Sigma^{-1}C)]^{\prime}~ \Sigma^{-1}
\end{eqnarray}
where $C = (c(t_0-t_1),...,c(t_0-t_n))'$ and $\Sigma$ is an $n
\times n$ matrix with $\it{(i, j)}$th element $c(t_i-t_j)$. The
Kriging variance can be written as
\begin{eqnarray}
\label{5'}
\sigma^2(t_0) = c(0) -C' \Sigma^{-1} C+(x
-X'\Sigma^{-1} C)' (X'\Sigma^{-1} X)^{-1} (x - X'\Sigma^{-1} C)
\end{eqnarray}
When $p=0$ and $f_0 (t)=1$, universal kriging reduce to ordinary
kriging.

 In universal kriging, the optimal value of
$\lambda$ (equation (\ref{5})) can be written as $\lambda_U =
\Sigma_U^{-1} C_U$ where $\lambda_U
=(\lambda_1,...,\lambda_n,-m_0,...,-m_p)'$ and $m_i s$ are
lagrange multipliers that insure $\lambda ^{\prime} X =
x^{\prime}$ and $C_U =(c(t_0
-t_1),...c(t_0-t_n),1,f_1(t_0),...,f_p(t_0))^{\prime}$. Then
kriging predictor at $t_0$ is
\begin{eqnarray}
\label{b1}
 \hat{Z}(t_0) ={\bf Z'}_U \Sigma_U ^{-1} C_U =V'_U C_U
\end{eqnarray}
where $V_U =\Sigma_U^{-1} {\bf Z}_U$ , ${\bf Z}_U
=(Z(t_1),...,Z(t_n),0,...,0)'$ which is an $(n+p+1)\times 1$
vector. In equation (\ref {b1}) by writing $V'_U =(V'_1 ,V'_2)$ so
that $V_1$ is $n \times 1$ and $V_2$ is $(p+1) \times 1$, then
 $V_U=\Sigma_U ^{-1} Z_U
 =\left[\begin{array}{cc}
  \Sigma & X \\
  X' & O
\end{array}\right]^{-1} \left[\begin{array}{c}
  Z \\
  0
\end{array}\right]=\left[\begin{array}{c}
  V_1 \\
  V_2
\end{array}\right]
$ or $\left[\begin{array}{cc}
  \Sigma & X \\
  X' & o
\end{array}\right]\left[\begin{array}{c}
  V_1 \\
  V_2
\end{array}\right]=\left[\begin{array}{c}
  Z \\
  0
\end{array}\right]$
and dual kriging equations is obtained as
\begin{eqnarray}
\label{12}
\left\{
\begin{array}{ll}
\Sigma V_1 +X V_2 = Z \\
 X'V_1= 0
 \end{array}
\right.
\end{eqnarray}
By solving this system and replacing in relation (\ref{3}),
 predictor of $Z(t_0)$ can be written as
\begin{eqnarray*}
\hat{Z} (t_0) =V'_1 C+V'_2{\bf x}
\end{eqnarray*}
\section{Spline}
Data {\bf Z} of random field Z(.) is given at locations $\{t_i
\in D \subset R^d , d>1\}$. Consider the problem of estimating
unknown function $\it{g}$ in the model
\begin{eqnarray}
\label{11} Z_i =g(t_i)+e_i ,\hspace{5mm} i=1,...,n
\end{eqnarray}
To fit $\it{g}$ properly, penalized sum of squares criterion is
defined as
\begin{eqnarray}
\label{b2} S(g,\lambda) = \Sigma_{i=1}^n (Z_i -g(t_i))^2 + \alpha
J_{r+1}^d(g)
\end{eqnarray}
where $\alpha > 0$ is smoothing parameter. A function $\hat g$
which minimizes penalized sum of squares criterion is called
Spline. The second term in equation (\ref {b2}) is
\begin{eqnarray*}
J_{r+1}^d(g) &=& \int|\nabla^{r+1}g(t)|^2 dt \cr
             &=& \Sigma_{|m|=r+1} {r+1\choose m}
\int(\frac{\partial^{r+1}g(t)} {\partial
t[1]^{m[1]},\dots,\partial t[d]^{m[d]}})^2 dt
\end{eqnarray*}
where $\nabla^{r+1}$ is $(r+1)$-fold iterated gradient of $g$, $t
=(t[1],\dots,t[d])$,
\begin{eqnarray*}
J_{r+1}^d (g)=\Sigma_{|m|=r+1} {r+1\choose m}   \int (\frac{\partial^{r+1} g(t)}
{\partial t[1]^{m[1]},\dots,\partial t[d]^{m[d]}})^2 dt
\end{eqnarray*}
where $m=(m[1],\dots ,m[d])$ and $|m|=m[1]+\dots+m[d]$.

For $d=2$, a function $\hat{g}$ which minimizes penalized sum of
squares (\ref {b2}) is called Thin Plate spline. To determine a
proper value of $\alpha$ can refer to 
Gu (2002), Hart (2005) and Hardle (2006).

Now finding dual equations for spline in case d=2 is considered
 (because dimension of our data is d=2.) Smoothing Spline of degree 2 is
\begin{eqnarray}
\label{13} \hat{Z} (t_0) =a_0 +a_1 x_0 +a_2 y_0 +\Sigma_{i=1}^n
b_i e(t_0 -t_i)
\end{eqnarray}
where
\begin{eqnarray*}
e({\bf h}) ={||{\bf h}||}^2\log({||{\bf h}||}^2)/16\pi
\end{eqnarray*}
In relation (\ref{13}), $a = (a_0,a_1,a_2)'$ and $b = (b_1,\dots,
b_n)'$ solve
\begin{eqnarray}
\label{15}
\left\{
\begin{array}{ll}
\label{15}
K_{\alpha} {\bf b} +X{\bf a} = Z \\
 X'{\bf b} =0
\end{array}
\right.
\end{eqnarray}
 where $K_{\alpha} = K + n \alpha I$ is an $n \times n$ matrix with $\it(i,j)$th
 elementary
$e(t_i - t_j)$, X is an $n \times 3$ matrix with $\it{i}$th row
$(1, x_i, y_i),~~ t_i = (x_i,y_i)'$ and $0\leq \alpha \leq \infty$
is the smoothing parameter.
\section{Application of Spline and Kriging to Prediction}
Dual equations (\ref{12}) and (\ref{15}) show that the form of
these equations for universal kriging and spline are the same,
just generalized covariance in Spline is used instead of
covariogram. In kriging method, when the second order stationary
condition does not satisfied or anyway the IRFk's is used,
generalized covariances are applied. Therefore dual equations of
kriging and spline methods are equal. Consequently methods of
kriging and spline are similar (theoretically), but they can be
different practically. In the next section these two methods are
compared in an epidemiological problem.
\subsection{Data Set and Practical Comparison}
Here data of taberculosis infection prevalence in the cities of
Iran on the year 1999 is considered. The random field is
nonstationary and data has a trend, therefore data is detrended
by median polishing. To estimate covariogram, Classic estimator
is applied and Gaussian model is chosen as the best model of
covariogram for this data set.

To compare the methods performances, a criterion should be
considered. Cross validation is a popular means of assessing
statistical estimation and prediction. If the variogram model
described adequately spatial dependencies implicit in data set,
then predicted value $\hat{Z} (t_0)$
 should be close to the true value $Z(t_0)$. Ideally additional
observations on $\it{Z(.)}$ to check this, or initially some of
the data might set aside to validate spatial predictor. More
likely, all of the data are used to fit the variogram, build the
spatial predictor, and there is no possibility of taking more
observations. In this case the cross validation approach can be
used. Let $2\gamma(h,\hat{\theta})$ be the fitted variogram model
(obtained from the data); now delete a datum $Z(t_j)$ and predict
it with $\hat{Z_{-j}} (t_j)$ [based on $2\gamma(h,\hat{\theta})$
 variogram estimator and the data ${\bf Z}$ without $Z(t_j)$]. Its
associated mean - square prediction error is $\sigma _{-j}
^2(t_j)$ which depends on the fitted variogram model.

The closeness of prediction values to the true values can be
characterized as the standardized Mean Square error of Prediction
\begin{eqnarray*}
\label{14} MSP=[1/n (\Sigma_{i=1}^n {\frac{Z(t_j) -
\hat{Z}_{-j}(t_j)}{\sigma_{-j} (t_j)}})^2]^{1/2}.
\end{eqnarray*}
In this paper, spline and kriging methods is compared by this
criterion and the better method which has smaller MSP is
determined. For this data set, gaussian model with nugget effect
equal to 39.8 is the best covariogram model to kriging
prediction. In spline method the smoothing parameter should be
determined and for this data set, the best value which minimizes
penalized sum of squares criterion equals $\alpha=208.6601$.

Cross validation criterion is applied to compare the methods.
Programs for computations is written in R and SPLUS environments
for the two dimensional data set.
 Cross validation criterion in kriging method is equal to
0.0239 and in spline method, it is equal to 0.0461.

Consequently, kriging method has a better performance than spline
for this data set. This result can be reasonable because in
spline usually a special generalized covariance function is used
but in kriging this function is characterized based on the data.
Therefore for some data sets, kriging method could have better
performance than spline.
\section {Conclusion}
Under certain conditions kriging and spline methods are
equivalent, but in practice there
 are differences between these methods. For instance in spline usually
 a particular generalized covariance function is used but in kriging,
this function is determined based on data, therefore it is
expected that kriging has a better performance in some situations.
In this paper these methods are applied to predict rate of
taberculosis infection prevalence which is a noticeable problem in
medicine.  The data has measured at two dimensional sites and
computations are carried out in R and SPLUS environments. For the
data set, computations show that kriging method has a better
performance than spline. Consequently application of Kriging can
be a preferable method of prediction.

\end{document}